# Determinants of Financial Performance of Microfinance Banks in Kenya

King'ori, S. Ngumo[1]   Kioko, W. Collins[2]   Shikumo, H. David[3]
1. MBA Student - University of Nairobi, Kenya
2. DBA Student - University of Nairobi, Kenya
3. PhD Finance Student - Jomo Kenyatta University of Agriculture and Technology, Kenya

**Abstract**
Microfinance provides strength to boost the economic activities of low-income earners and thus contributes to eradication of poverty. However, microfinance institutions face stringent competition from commercial banks; the growth of microloan activities of commercial banks may confront microfinance institutions with increased competition for borrowers. In Kenya, the micro finance sector has extremely high competition indicated by the shifting market share and profitability. This study sought to examine the determinants of financial performance of Microfinance banks in Kenya. The study adopted a descriptive research design and used secondary data from 7 Microfinance banks for a period of 5 years from 2011 to 2015. The data collected was analyzed using correlation and regression analysis. The study found a positive and statistically significant relationship between operational efficiency, capital adequacy, firm size and financial performance of microfinance banks in Kenya. However, the study found an insignificant negative relationship between liquidity risk, credit risk and financial performance of microfinance banks in Kenya. The study concluded that there is direct relationship between operational efficiency, capital adequacy, firm size and financial performance of microfinance banks in Kenya.

**1.0 INTRODUCTION**
*1.1 Background of the Study*
Micro-finance is one of the ways of building the capacities of the poor who are largely ignored by commercial banks and other lending institutions and graduating them to sustainable self-employment activities by providing them financial services like credit, savings and insurance (Anand & Kanwal, 2011). Microfinance is associated with programs that benefit clients with serious subsistence problems in developing countries. For many years, microfinance overlapped with microcredit (small loans) often without traditional guarantees, aimed at improving the lives of clients and their families or at sustaining small-scale economic activities (Addisalem, 2015). The Microfinance sector has demonstrated that poor people are viable customers as long as their financing is approached in the right way, which required that moral hazard, adverse selection and other agency problems are mitigated (Dokulilova, Janda & Zetek, 2009).

Microfinance provides strength to boost the economic activities of low-income group people and thus contributes to eradication of their poverty (Almas & Mukhtar, 2014). However, MFIs are confronted with the microfinance schism, which covers two fundamental requirements. The social requirement of poverty alleviation (social performance) carried by the welfarists, and the economic requirement of profitability and viability of institutions (financial performance) defended by the institutionalists (Berguiga, 2008). The institutionist school thought financial deepening is the main aim of microfinance. That is, the setting up of a separate system of sustainable financial intermediation for the poor who are either neglected or are underserved by the formal financial system. On the other hand, the welfarist approach focuses on depth (number of clients reached) rather than breadth of outreach (poverty level of clients) and accept subsidies on an ongoing basis (Yenesew, 2014).

The Kenyan microfinance sector is one of the most vibrant in Sub-Saharan Africa. It includes a diversity of institutional forms and a large branch network to serve the poor (FSD Kenya, 2012). The microfinance act (2006) and the microfinance regulations (2008) set out the legal, regulatory and supervisory framework for the microfinance industry in Kenya (Association of Microfinance Institutions, 2013). The need for microfinance in Kenya has been driven by a series of interrelated constraints on the development of a banking and finance sector. These key constraints have been the structure and composition of the Kenyan banking and finance sector; a lack of the appropriate regulation and governance required for quality improvements in banking and finance and the conservative commercial business practices of profit focused banking institutions (Alastair, 2015).

*1.1.1 Financial Performance*
Financial performance is the measure of organizations achievement on the goals, policies and operations stipulated in monetary terms. It involves the financial health and can be compared between similar firms in the same industry (Agola, 2014). Financial performance of a company, being one of the major characteristics, defines competitiveness, potentials of the business and economic interests of the company's management and reliability of present or future contractors (Dufera, 2010). In the MFI context, financial performance is the ability of a MFI to keep on going towards microfinance objective without donor support (Thapa, 2008). The main aim of every micro-finance institution is to have operations that are profitable in order to maintain stability and improve on sustainability and growth (Agola, 2014). Thus, Microfinance Institutions (MFIs) should seek to maximize performance in many areas, whether it is social or economical (Jørgensen, 2011).

Financial performance analysis is an appraisal of the feasibility, solidity and fertility of a business, sub-business or mission (Bhunia, Mukhuti & Gautam, 2011). Good financial performance rewards the shareholders for their investment (Ongore & Gemechu, 2013). A firm's financial performance, in the view of the shareholder, is measured by how better off the shareholder is at the end of a period, than he was at the beginning and this can be determined using ratios derived from financial statements; mainly the balance sheet and income statement, or using data on stock market prices (Baraza, 2014). Financial performance can be measured through various financial measures such as profit after tax, return on assets (ROA), return on equity (ROE), earnings per share and any market value ration that is generally accepted (Yenesew, 2014). The return on assets ratio (ROA) is an important financial performance ratio because it measures the efficiency with which the company is managing its investment in assets and using them to generate profit (Jørgensen, 2011).

### 1.1.2 Factors Influencing Financial Performance of Microfinance Banks

A good performing microfinance industry is vital in sustaining the stability of the micro banking system. Poor financial performance deteriorates the capacity of MFIs to absorb negative shocks, which subsequently affect solvency (Yenesew, 2014). Better financial performance leads the lenders to recover full cost or make profit, and building institutions that can sustain themselves for a considerable period without continued reliance on government subsidies or donor funds. MFIs financial performance is based on the extent to which service users directly pay the full cost of providing services (Adhikary, 2014). As such operational efficiency, capital levels, liquidity risk, credit risk and size are some of the major factors that influence financial performance of microfinance banks.

Operational efficiency refers to the ability of a microfinance program to deliver a specific service with minimum costs (Adhikary, 2014). Operational efficiency is performance measure that shows how well MFIs is streamlining its operations and takes in to account the cost of the input and/or the price of output (Ongore & Gemechu, 2013). Efficiency in expense management should ensure a more effective use of MFIs loanable resources, which may enhance MFIs profitability. Inefficiency is one of the significant risk factors for sustainable microfinance as large numbers of institutions are still far from minimal scale or the efficiency required to cover costs. Operational efficiency is usually measured using operating efficiency ratio (OER) where lower OER is preferred over higher OER as lower OER indicates that operating expenses are lower than operating revenues (Dufera, 2010).

Capital is the amount of own fund available to support the bank's business and act as a buffer in case of adverse situation. Capital creates liquidity for a financial institution since deposits are essentially other people's money, which can be recalled at any time (Dang, 2011). In the event of loss of assets, higher capital level relative to its assets ensures the institutions would have sufficient funds of its own to cover the loss or there is sufficient level of capital required to absorb potential losses while providing financial sustainability (Adhikary, 2014). As such, in the presence of asymmetric information, a well-capitalized bank could provide a signal to the market that a better-than-average performance should be expected (Kahiga, 2014). Capital level/adequacy is normally proxied using proportion of MFI equity to total assets.

Liquidity refers to the ability of institutions to meet demands for funds. Liquidity risk arises when a microfinance bank is unable to meet its cash requirements or payment obligations timely and in a cost-efficient manner (Idama et al., 2014). MFI with inadequate liquidity might be less immune towards future uncertainty, timely delay of refinancing, disruption in meeting growth projections and increased portfolio at risk (Brom, 2009). To reduce liquidity risk, each microfinance bank branch needs to prepare a daily fund plan that guides the matching of cash inflows from loan repayment and saving deposits with cash outflows for the branch on a daily basis (Idama et al., 2014). Loan to total assets ratio (LAR) is normally used to measure the liquidity position of MFI that indicates the percentage of total assets used to provide the loan (Adhikary, 2014).

Credit risk is the financial loss that a lender will suffer because of a borrower's failure to perform according to the terms and conditions of the credit or loan agreement. Effective management of credit risk through proper management results in the improvement of earnings and reduces insolvency (Sule, 2012). Credit risk is not confined to a microfinance banks' loan portfolio alone, but can also exist in its other assets and activities. Credit risk affects the profitability and the general performance of any financial institution and is one of the major risks to microfinance banks sustainability. Thus, managing credit risk is an integral part of microfinance bank operating techniques, with reducing the risks requiring a major operational effort (Idama et al., 2014).

The size of an institution plays an important role in determining the kind of relationship the firm enjoys within and outside its operating environment and hence profitability. The modern intermediation theory predicts efficiency gains related to size of a financial institution, owing to economies of scale (Kahiga, 2014). Smaller MFIs in particular are at a disadvantage, struggling to cover the industry's high operational costs and diversify their products in order to compete with larger microfinance providers (Muriu, 2011). In addition, large firms are more diversified than small ones and have greater market power and during good times may have relatively more organizational slack (Addisalem, 2015). Size captures the economies or diseconomies of scale of an institution and normally the natural logarithm of total asset of MFIs is used as a proxy of size (Cull et al., 2007).

**1.1.3 Microfinance Banks in Kenya**
The Microfinance Act (2006) defines Microfinance bank or a deposit-taking microfinance business as a business in which the person conducting the business holds himself out as accepting deposits on a day-to-day basis. Microfinance banks are registered under the Microfinance Act (2006) and are not fully registered banks but are subject to many of the same conditions under the prudential control of the Central Bank, given that they use customer deposits to raise capital for independent loans (Alastair, 2015). Microfinance banks accept demand deposits and use the deposits as a means to generate capital for the extension of credit to customers (Alastair, 2015).In the microfinance operations, Kenya is ranked first in Africa and fifth in the world, respectively (Ayele, 2014).

The Kenya's microfinance sector comprises of nearly 250 MFIs, with only 56 of these being registered with the Association of Microfinance Institutions, an umbrella body. In Kenya as at December 2015 there were12 deposits taking microfinance institutions. Among the major players in the sector, include Faulu Kenya, Kenya Women Finance Trust (KWFT), Small and Medium Enterprise Programme (SMEP), Rafiki Microfinance Bank, Century MFI, Sumac MFI bank limited, Uwezo MFI amongst others (Njenje & Bengi, 2016). Kenya's Micro finance industry focuses on delivering financial services to low-income individuals and Micro and Small Enterprises (MSEs) engaged in non-farm productive activities. Over time, MFIs have introduced significant innovations in products and services, which are patronized by MSEs (Gibson, 2012). The total assets of the microfinance sector registered a stable growth over the past three years with the sector being dominated by banks (Agola, 2014).

*1.2 Research Problem*
The microfinance industry has been growing at a significant rate in several countries and it has become an important sub-sector of the formal financial markets (Assefa, Hermes & Meesters, 2010). However, the microfinance industry, along with all the players in it, is quickly changing (Yenesew, 2014).The number of microfinance service providers has also increased considerably and with the growth of the industry and the saturation of markets, increased competition has been documented in many countries (Porteous, 2006). Many microfinance institutions have secured high loan repayment rates, but, so far, relatively few earn profits (Cull, Demirgüç-Kunt & Morduch, 2007). MFIs also face stringent competition from commercial banks; the growth of microloan activities of commercial banks may confront MFIs with increased competition for borrowers (Addisalem, 2015).

In Kenya, the microfinance sector has experienced extremely high competition evidenced by the shifting market share and profitability. The competition is among the MFIs sector, mainstream commercial banks and the telecommunication money transfer platforms such as Mpesa (Okombo, 2015). According to AMFI (2013), while over the time credit-only institutions have been slowly improving, banks and DTM improved in 2010-2011 but then worsened slightly in 2011-2012. As such, Microfinance banks in Kenya have also reported very high competitive pressure in terms of pricing since they have less flexibility to adjust prices due to their financial structure (IMFI, 2013). Thus, the need to investigate the determinants of financial performance of Microfinance banks in Kenya.

Various scholars have also examined the various factors that influence the financial performance of microfinance institutions globally and locally. A study by Hien (2009) found that regulatory status has no direct effect on financial sustainability and outreach. However, the study focused on social performance of MFIs measured in terms of sustainability and outreach as opposed to financial performance. A study by Biwott and Muturi (2014) found that number of borrowers, capital adequacy and branch network had the greatest influence on the performance of microfinance institutions. Kimando, Kihoro and Njogu (2012) established that financial regulations, number of clients served, financial coverage and volume of credit transacted highly affected the sustainability of microfinance institutions. However, the above studies investigated MFIs in specific geographical areas in Kenya and focused on both deposit taking MFIs and credit only MFI. In addition, most studies on microfinance institutions focus more on outreach, financial and operational sustainability on all types of MFIs. Thus, the need to investigate; which are the determinants of financial performance of deposit taking microfinance institutions in Kenya?

*1.3 Research Objective*
To investigate the determinants of financial performance of Microfinance banks in Kenya

**2.0 LITERATURE REVIEW**
A study by Arthur et al. (2013) examined the degree of financial performance in selected Microfinance institutions in central region, Uganda. The study employed the ex-post facto or retrospective, prospective designs and descriptive survey design, and specifically descriptive comparative and descriptive correlation strategies. The findings of the study revealed that the degree of financial performance in the microfinance institutions in central region Uganda was high. Mwizarubi, Singh and Mnzava (2015) also examined the impact of modern Microfinance Institutions (MFIs) capital structure variables on MFI's financial sustainability. The study found that deposit mobilization is the most crucial determinant of financial sustainability amongst other MFI capital structure variables, followed by shareholders' equity, debt (commercial borrowing) and lastly going public.

In their study, Tilahun and Dereje (2012) assessed the financial performance of Ethiopian MFIs using a descriptive research design. The findings of the study revealed that there was a negative shift in the performance indicators with the decline of the gross loan portfolio. The study also found that the portfolio at risk rose during 2008 and 2009 indicating deterioration of portfolio quality while the number of active borrowers (outreach) declined though there was an increase in number of staff members in the MFI. Muriu (2011) explored the impact of financing choice on microfinance profitability using an unbalanced panel dataset comprising of 210 MFIs across 31 countries operating from 1997 to 2008. The study found a proportionally higher deposit as a ratio of total assets was associated with improved profitability. Further, the study revealed that MFIs with a higher portfolio-assets ratio are more profitable but the impact depends on MFI age.

A study by Jørgensen (2011) examined the factors that determine profitability of microfinance institutions using a sample of 879 MFIs. The study findings established that the factors that statistically influenced profitability positively were the capital asset ratio, age (new) and the gross loan portfolio. The study also found that operating expense over loan portfolio had a positive influence but the number of active borrowers had a negative influence. Assefa, Hermes and Meesters (2010) investigated the relationship between competition and the performance of microfinance institutions (MFIs). The findings of the study revealed that competition among MFIs was negatively associated with various measures of performance.

Cull, Demirgüç-Kunt and Morduch (2009) examined the implications of regulation and supervision on MFIs profitability and their outreach to small-scale borrowers and women. The study findings established that supervision is negatively associated with profitability. In addition, the study revealed that supervision is associated with substantially larger average loan sizes and less lending to women although it is not significantly associated with profitability. Narwal, Pathneja and Yadav (2014) studied the performance variables of banking sector and microfinance institutions in India over a study period of six years 2006 to 2012. The findings of the study established that performance of microfinance institutions mainly rotate around two variables, which include size and spread to total assets.

In Kenya, Njenje and Bengi (2016) assessed the financial factors that affect the growth of microfinance institutions (MFIs) in Bahati Sub-county. The study findings established that there exists a strong, positive and statistically significant relationship between financial literacy and growth of MFIs; whereas relationship between interest rates and growth of MFIs was not statistically significant. Njeru et al. (2015) explored the effect of loan repayment on financial performance of deposit taking SACCOs in Mount Kenya Region. The study findings established that there was positive relationship between loan repayment and financial performance of deposit taking SACCOs in Mount Kenya Region.

Okombo (2014) examined the impact of low transactional costs on the financial performance of deposit taking MFIs. The study findings established that there was positive and statistically significant relationship between the low transaction costs and financial performance. Agola (2014) explored the relationship between credit policy and financial performance of microfinance institutions in Kenya. The findings of the study revealed a positive relationship between financial performance, credit policy, credit risk controls, credit appraisal and collection policy. Ongaki (2012) examined the determinants of profitability of deposit-taking microfinance institutions and co-operative societies. The study findings established that there was a positive relationship between profit ratio and interest income ratio and non-interest income ratio. The study also found that there was negative relationship between profit ratio and noninterest expense ratio and liquidity ratio.

A study by Gisbson (2012) explored the factors that determine the operational sustainability of micro finance institutions in Kenya. The findings of the study established that the factors affecting operations and financial sustainability are capital/ asset ratio and operating expenses/loan portfolio. Baraza (2014) also investigated the relationship between funding structure and financial performance of Microfinance institutions in Kenya. The study found that debt to equity ratio had a negative correlation with financial performance meaning the more debt a firm employ in financing its operations the inferior financial performance it registers. The study also found that deposits to assets ratio had a positive correlation with financial performance implying that the more deposits a microfinance institution accepts the higher the financial performance.

## 3.0 METHODOLOGY
This study sought to establish the determinants of financial performance of Microfinance banks in Kenya. Therefore, the study employed a descriptive research design. The population of the study was made of 7 microfinance banks, which had been in operation since 2011. The study used secondary data from the 7 Microfinance banks in Kenya as at 31/12/2005. The data was extracted from the Microfinance banks financial reports i.e. the statement of comprehensive income and financial position for a period of 5 years from the years 2011 – 2015. The data collected was analyzed using descriptive and inferential statistics with the help of the Statistical Package for Social Studies. Descriptive statistics summarized the data using the mean and standard deviation while correlation and regression analysis were used to determine relationship between the dependent and the independent variables. The regression was as follows

$$ROA = \beta_o + \beta_1(OER) + \beta_2(CA) + \beta_3(LAR) + \beta_4(CR) + \beta_5(FS) + \varepsilon$$

Where,

$ROA$ = Return on Assets used as a proxy for financial performance
$OER$ = Operation Efficiency Ratio used as a proxy for operating efficiency
$CA$ = Capital Adequacy proxied using the proportion of MFI equity to total assets
$LAR$ = Loan to Asset Ratio used as a proxy for liquidity position
$CR$ = Credit risk proxied using the non-performing loans ratio (NPLR)
$FS$ = Firm size proxied using natural log total assets

## 4.0 RESEARCH FINDINGS AND DISCUSSION
### 4.1 Descriptive Statistics
**Table 4.1 Descriptive Statistics**

|     | N  | Minimum | Maximum | Mean    | Std. Deviation |
|-----|----|---------|---------|---------|----------------|
| ROA | 35 | -.269   | .053    | -.02105 | .067424        |
| OER | 35 | .027    | 2.172   | .25271  | .341111        |
| CA  | 35 | .100    | 1.410   | .43080  | .321741        |
| LAR | 35 | .210    | 2.980   | .54334  | .592279        |
| CR  | 35 | .017    | .889    | .12617  | .179002        |
| FS  | 35 | 4.078   | 10.383  | 7.26603 | 2.008333       |

The results on table 4.1 indicate that the average ROA (return on assets) for the microfinance banks during the study periodic -0.0211, which is an indication that the average financial performance of microfinance banks between 2011 and 2015 in Kenya was negative. The results also indicate that the average OER (operation efficiency ratio) for microfinance banks is 0.25271 while the average CA (capital adequacy ratio) is 0.4308, which indicate that the average capital adequacy level for microfinance banks in Kenya from 2011 to 2015 was 43.08%. Further, the results show that the average LAR (loan to asset ratio) is 0.5433, which indicates that the average liquidity risk for microfinance banks in Kenya from 2011 to 2015 was very high. The results indicate that the average CR (credit risk) is 0.1261, which indicates that low credit risk for the microfinance banks from 2011 to 2015. Finally, the results show that the average size of the microfinance banks was 7.266 for the period between 2011 and 2015.

### 4.2 Correlation Analysis
**Table 4.2 Correlation Analysis**

|     | ROA     | OER   | CA      | LAR    | CR     | FS |
|-----|---------|-------|---------|--------|--------|----|
| ROA | 1       |       |         |        |        |    |
| OER | .070    | 1     |         |        |        |    |
| CA  | -.323   | -.066 | 1       |        |        |    |
| LAR | -.142   | -.079 | .678**  | 1      |        |    |
| CR  | -.278   | -.007 | -.030   | -.043  | 1      |    |
| FS  | .511**  | -.208 | -.641** | -.339* | -.407* | 1  |

**. Correlation is significant at the 0.01 level (2-tailed).
*. Correlation is significant at the 0.05 level (2-tailed).

The results on table 4.2 indicate that there is a weak positive correlation between financial performance and operational efficiency but a strong positive correlation between financial performance and firm size. The results also indicate that there is a weak negative correlation between capital adequacy, liquidity risk, credit risk and financial performance of microfinance banks in Kenya.

*4.3 Regression Analysis*
**Table 4.3 Summary of the Regression Results**

|  | B | Std. Error | T | Sig. |
|---|---|---|---|---|
| (Constant) | -.322 | .174 | -1.850 | .075 |
| OER | .031 | .008 | 3.796 | .001 |
| CA | .064 | .030 | 2.107 | .044 |
| LAR | -.020 | .013 | -1.474 | .151 |
| CR | .016 | .027 | .568 | .574 |
| FS | .005 | .002 | 2.505 | .018 |
| F-value (5,29) | 6.347 | P-value | .000 | |
| R | .723 | R-Square | .522 | |
| Adjusted R Square | .440 | | | |

Dependent Variable: ROA

The results on table 4.3 indicate that the R-square value is 0.522, which indicates that the independent variables (operational efficiency, capital adequacy, liquidity risk, credit risk and firm size) explain 52.2% of the variation in the dependent variable (financial performance). Additionally, the results indicate that the F statistics value is 6.347 with a p value of 0.000 which is less that the significance value of 0.05 (0.000<0.05) hence an indication that the regression model is significant and a good predictor of the relationship between the independent variables and the dependent variable.

The results also indicate that there is a statistically positive and significant relationship between operational efficiency (OAR), capital adequacy (CA), firm size (FS) and financial performance of microfinance banks in Kenya. This means that there is a direct relationship between operational efficiency, capital adequacy, firm size and financial performance of microfinance banks in Kenya. The findings are similar to that of Yenesew (2014) who found that operational efficiency and size of MFIs affect MFIs financial performance significantly. Sudharika and Madurapperuma (2016) also found that operational efficiency ratio and capital adequacy ratio affect MFIs financial performance significantly. Narwal, Pathneja and Yadav (2014) also found that performance of MFIs rotates around mainly size and spread to total assets. However, the results show an insignificant negative relationship between liquidity risk (LAR), credit risk (CR) and financial performance of microfinance banks in Kenya. As such, Ongaki (2012) found a negative relationship between liquidity ratio and profitability of deposit-taking microfinance institutions in Kenya

**5.0 CONCLUSION**
The findings of the study established that operational efficiency, capital adequacy and firm size significantly and positively influences the financial performance of microfinance banks in Kenya. The study therefore concludes that there is direct relationship between operational efficiency, capital adequacy, firm size and financial performance of microfinance banks in Kenya. The study further found that liquidity risk and credit risk do not have statistically significant relationship with financial performance of microfinance banks of the Kenya. This leads to the conclusion that liquidity risk and credit risk do not affect the financial performance of microfinance banks in Kenya since there is a minimum liquidity ratio requirement set by the central bank of Kenya and microfinance banks in Kenya have low credit risk levels.